\documentstyle[titlepage,preprint,aps,floats,epsfig]{revtex}
\newcommand{\beq}{\begin{equation}}
\newcommand{\eeq}{\end{equation}}
\newcommand{\eq}[1]{eq.(\ref{#1})}
\begin{document}
\draft
\preprint{UK/03-01}
\tighten
\title {Three-Loop Radiative-Recoil
Corrections to Hyperfine Splitting in Muonium}
\medskip
\author {Michael I. Eides \thanks{E-mail address:
eides@pa.uky.edu, eides@thd.pnpi.spb.ru}}
\address{Department of Physics and Astronomy,
University of Kentucky,
Lexington, KY 40506, USA\\
and
Petersburg Nuclear Physics Institute,
Gatchina, St.Petersburg 188300, Russia}
\author{Howard Grotch\thanks{E-mail address: asdean@pop.uky.edu}}
\address{Department of Physics and Astronomy, University of Kentucky,
Lexington, KY 40506, USA}
\author{Valery A. Shelyuto \thanks{E-mail address:
shelyuto@vniim.ru}}
\address{D. I.  Mendeleev Institute of Metrology,
St.Petersburg 198005, Russia}

\maketitle

\begin{abstract}
We calculate three-loop radiative-recoil corrections to hyperfine
splitting in muonium generated by the diagrams with the first order
electron and muon polarization loop insertions in graphs with two
exchanged photons. These corrections are enhanced by the large
logarithm of the electron-muon mass ratio. The leading logarithm
squared contribution was obtained a long time ago. Here we calculate
the single-logarithmic and nonlogarithmic contributions. We previously
calculated the three-loop radiative-recoil corrections generated by
two-loop polarization insertions in the exchanged photons. The current
paper therefore concludes calculation of all three-loop
radiative-recoil corrections to hyperfine splitting in muonium
generated by diagrams with closed fermion loop insertions in
the exchanged photons. The new results obtained here improve the
theory of hyperfine splitting, and affect the value of the
electron-muon mass ratio extracted from experimental data on the
muonium hyperfine splitting.
\end{abstract}


\section{Introduction}

Recently we initiated a program of calculating of all three-loop
radiative-recoil corrections to hyperfine splitting (HFS) in muonium
\cite{egs02}.  Three-loop radiative-recoil corrections are enhanced by
the presence of the cube of the large logarithm of the electron-muon
mass ratio \cite{es0}.  All leading logarithm cubed and logarithm
squared contributions of this order were calculated a long time ago
\cite{es0,eks89,kes90} (see also reviews in \cite{egs02,egs01r}). As
the first step of our program we obtained in \cite{egs02}  previously
unknown single-logarithmic and nonlogarithmic radiative-recoil
corrections of order $\alpha^2(Z\alpha)(m/M)\widetilde E_F$\footnote{We
define the Fermi energy  as

\beq      \label{baremuonfermi}
\widetilde{E}_{F}=\frac{16}{3}Z^4\alpha^2
\frac{m}{M} \left(\frac{m_r}{m}\right)^{3}ch\:R_{\infty},
\eeq

\noindent
where $m$ and $M$ are the electron and muon masses, $\alpha$ is the
fine structure constant, $c$ is the velocity of light, $h$ is the
Planck constant, $R_{\infty}$ is the Rydberg constant, and $Z$ is the
nucleus charge in terms of the electron charge ($Z=1$ for muonium).
The Fermi energy  $\widetilde{E}_{F}$ does not include the muon
anomalous magnetic moment $a_\mu$ which does not factorize in the case
of recoil corrections, and should be considered on the same grounds as
other corrections to hyperfine splitting.} generated by graphs with
two-loop polarization insertions (irreducible and reducible) in the
two-photon exchange diagrams. As the next logical stage in
implementing our program, we present below the calcuulation of all
single-logarithmic and nonlogarithmic three-loop radiative-recoil
corrections generated by diagrams with one-loop electron and muon
polarization insertions in the exchanged photons. There are four gauge
invariant sets of such three-loop diagrams, and we calculate all their
contributions.

\section{Radiative-Recoil Correction of Order
$\alpha(Z\alpha)(\lowercase{m}/M)\widetilde E_F$}
\label{radinsel}

All four sets of diagrams considered below can be obtained from the
two-photon exchange diagrams with the radiative photons in the electron
or muon lines by insertions of the one-loop electron or
muon polarization operators. As was discussed in \cite{egs02}, it is
sufficient to calculate contributions of these diagrams in the
scattering approximation. In the calculations below we use  the
approach developed earlier for analytic calculation of the two-loop
radiative-recoil corrections of orders $\alpha(Z\alpha)(m/M)\widetilde
E_F$ and $(Z^2\alpha)(Z\alpha)(m/M)\widetilde E_F$ in \cite{eks5,eks6}
(these corrections were also calculated numerically in \cite{sty}).  To
make this paper self-contained we first briefly remind the reader
of the main steps in calculation of the corrections induced by the
radiative photon insertions in the electron line.

\begin{figure}[ht]
\centerline{\epsfig{file=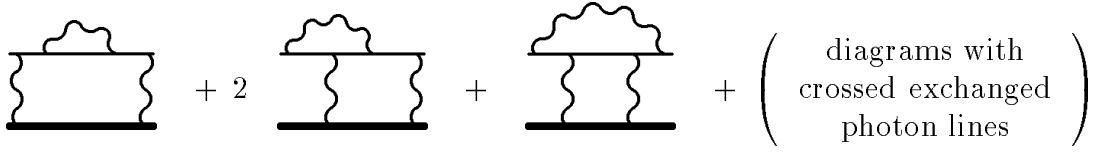,height=2cm}}
\vspace{0.5cm}
\caption{Electron-line radiative-recoil corrections}
\label{ellineradreclamb}
\end{figure}

The integral representation for the radiative
corrections of order $\alpha(Z\alpha)(m/M)\widetilde E_F$ generated by
the graphs with radiative insertions in the electron line in Fig.\
\ref{ellineradreclamb} has the form\cite{beks,egs98}

\begin{equation}      \label{electrcontr}
\delta E^{\,e-line} ~=~ \alpha (Z\alpha) \widetilde E_F~
\frac{1}{8 \pi^2 \mu} ~ \int_0^1 {dx} \int_0^x {dy} \int \frac{d^4 k}{i
\pi^2(k^2 + i0)^2}~
\end{equation}
\[
\biggl( \frac{1}{k^2
+ \mu^{-1}k_0 + i0} ~+~ \frac{1}{k^2 - \mu^{-1}k_0 + i0} \biggr)
\]
\[
\Bigl\{~ (~3k^2_0 ~-~ 2{\bf k}^2 ~)
\Bigl[~\frac{c_{1}  {\bf k}^2 ~+~ c_{2}  (k^2)^2}
{(~-k^2 ~+~ 2bk_0 ~+~ a^2~)^3}
~+~\frac{ c_{3}  k^2 ~+~ c_{4}  2k_0}
{(~-k^2 ~+~ 2bk_0 ~+~ a^2~)^2} \Bigr]
\]
\[
~-~ 3k_0  \Bigl[
~\frac{c_{5}  k^2 ~+~ c_{6}  k^2  2k_0}
{(~-k^2 ~+~ 2bk_0 ~+~ a^2~)^2}
~+~ \frac{c_{7}  k^2}
{~-k^2 ~+~ 2bk_0 ~+~ a^2~} \Bigr] \Bigr\}\equiv\sum_1^7\delta E^{e-line}_i,
\]

\noindent
where the dimensionless integration momentum $k$ is measured in
units of the electron mass, the small parameter $\mu$ is defined as
half the ratio of the electron and muon masses ($\mu=m/(2M))$,  the
auxiliary functions of the Feynman parameters $a(x,y)$ and $b(x,y)$ are
defined by the relationships

\begin{equation}
a^2 ~=~ \frac{x^2}{y(1 - y)}, ~~~~~~ b ~=~ \frac{1 - x}{1 - y},
\end{equation}

\noindent
and explicit expressions for the coefficient functions $c_i$ are collected
in Table \ref{table1}. We used the Yennie gauge for the radiative
photons in the derivation of the explicit expressions for these
coefficient functions $c_i$.

After the Wick rotation and transition to four-dimensional
spherical coordinates, the expression in \eq{electrcontr} acquires the
form

\beq   \label{generalprev}
\delta E^{e-line} ~~=~~  \frac{\alpha (Z\alpha)}{\pi^2}\frac{m}{M}
\widetilde E_F~\frac{1}{8\pi \mu^2} ~
\int_0^{\pi} {d \theta}~ \sin^2 {\theta}
~ \int_0^{\infty} {d k^2}~  ~{\cal D}(k, \theta)
\eeq
\[
\frac{k^2  }{k^2 + \mu^{-2} ~\cos^2 {\theta}}
~ \frac{1}{~(k^2 + a^2)^2 ~+~ 4 b^2 k^2 \cos^2{\theta}~}~~,
\]

\noindent
where

\beq  \label{difop}
{\cal D}(k, \theta) = \int_0^1 {dx} \int_0^x {dy}
\Bigl\{(2 + \cos^2{\theta})  \Bigl[(c_1 \sin^2{\theta} + c_2 k^2)
 \Bigl(\frac{\partial}{\partial a^2}\Bigr)^2 +
2 c_3  \frac{\partial}{\partial a^2} \Bigr] (k^2 + a^2)
\eeq
\[
- 8 b c_4 (2 + \cos^2{\theta}) \cos^2{\theta}
 \frac{\partial}{\partial a^2} +
12 b \cos^2{\theta} \Bigl(c_5  \frac{\partial}{\partial a^2}
-c_7 \Bigr)
- 12 c_6  \cos^2{\theta}  \frac{\partial}{\partial a^2}
(k^2 + a^2) \Bigr\}.
\]

\noindent
We introduced derivatives with respect to $a^2$ in order to reduce the
powers in the denominators before integration over angles.

Next we separate the $\mu$-dependent and $\mu$-independent terms in the
integrand with the help of the identity

\beq
\frac{1}
{\Bigr(~k^2 + \mu^{-2} ~\cos^2 {\theta} ~ \Bigl)
\Bigl[~(k^2 + a^2)^2 ~+~ 4~ b^2~ k^2 ~\cos^2{\theta}~\Bigr]}
\eeq
\[
=\frac{\mu^2}{~(k^2 + a^2)^2 ~-~ 4~ \mu^2~ b^2~ k^4~}
\left[\frac{1}{\mu^2k^2 +
\cos^2 {\theta}} ~-~ \frac{4 b^2 k^2}{(k^2 + a^2)^2 ~+~ 4 b^2 k^2
\cos^2{\theta}}\right]
\]
\[
\approx\frac{\mu^2}{~(k^2 + a^2)^2}
\left[\frac{1}{\mu^2k^2 +
\cos^2 {\theta}} ~-~ \frac{4 b^2 k^2}{(k^2 + a^2)^2 ~+~ 4 b^2 k^2
\cos^2{\theta}}\right],
\]

\noindent
where, at the last step, we omitted the term proportional to $\mu^2$
in the denominator before the square bracket since, as can be
shown, this term generates recoil corrections which are at least
quadratic in the recoil factor $\mu$. Then the integral for the
radiative corrections in \eq{generalprev} becomes a sum of
$\mu$-dependent and $\mu$-independent integrals

\beq   \label{generalprevsep}
\delta E^{\,e-line} ~~\approx~~  \frac{\alpha
(Z\alpha)}{\pi^2}\frac{m}{M} \widetilde E_F~\frac{1}{8\pi} ~
\int_0^{\pi} {d \theta}~ \sin^2 {\theta}
~ \int_0^{\infty} {d k^2}~  ~{\cal D}(k, \theta)
\eeq
\[
\frac{k^2 }{~(k^2 + a^2)^2}
\left[\frac{1}{\mu^2k^2 +
\cos^2 {\theta}} ~-~ \frac{4 b^2 k^2}{(k^2 + a^2)^2 ~+~ 4 b^2 k^2
\cos^2{\theta}}\right]=\delta E^{e-line}_{\mu i}+\delta E^{e-line}_{c
i}, \]

\noindent
which we will further call $\mu$-integrals and $c$-integrals,
respectively.

It is convenient to write the angular integrals in
\eq{generalprevsep} in terms of the standard functions $\Phi_n^i$
defined as

\beq
\Phi_n (k)\equiv \frac{1}{\pi \mu^2}
\int_0^{\pi} {d \theta} \sin^2 {\theta}   \cos^{2n} {\theta}
\frac{(k^2 + a^2)^2 - 4 \mu^2 b^2 k^4}
{\Bigr(k^2 + \mu^{-2} \cos^2 {\theta}  \Bigl)
\Bigl[(k^2 + a^2)^2 + 4 b^2 k^2 \cos^2{\theta}\Bigr]}
\eeq
\[
=\frac{1}{\pi}
\int_0^{\pi} {d \theta} \sin^2 {\theta}   \cos^{2n} {\theta}
\left[\frac{1}{\mu^2k^2 +
\cos^2 {\theta}} - \frac{4 b^2 k^2}{(k^2 + a^2)^2 + 4 b^2 k^2
\cos^2{\theta}}\right]
\]
\[
=\Phi_n^S (k) +\Phi_n^{\mu} (k) +\Phi_n^C (k).
\]

\noindent
Explicit expressions for these functions and their properties are
collected in Appendix \ref{appa} (see also \cite{egs98}).

The $c$-integrals with the functions  $\Phi_n^C (k)$ do not contain any
free parameters and generate only contributions linear in $m/M$ to HFS.
These integrals are pure numbers, which admit analytic calculation.
The $\mu$-integrals with the functions $\Phi_n^S (k)$, $\Phi_n^{\mu}
(k)$ parametrically depend on $\mu$ and generate both nonrecoil and
recoil contributions to HFS. Contributions  of a fixed order in the
small mass ratio (which are often enhanced by the large logarithms of
the mass ratio) can be extracted from the $\mu$-integrals with the help
of an auxiliary parameter $\sigma$ chosen such that it satisfies the
inequality $1\ll \sigma\ll \mu^{-1}$. The parameter $\sigma$ is used to
separate the momentum integration into two regions, a region of small
momenta $0\leq k\leq\sigma$, and a region of large momenta $\sigma\leq
k<\infty$. In the region of small momenta one uses the condition $\mu
k\ll 1$ to simplify the integrand, and in the region of large
momenta the same goal is achieved with the help of the condition $k\gg
1$. Note that for $k\simeq \sigma$ both conditions on the integration
momenta are valid simultaneously, so in the sum of the low-momenta and
high-momenta integrals all $\sigma$-dependent terms cancel and one
obtains a $\sigma$-independent result for the total momentum integral
(for a more detailed discussion of this method, see, e.g.,
\cite{eksann1}).  All contributions, up to and including the corrections
quadratic in the small mass ratio $m/M$, were analytically calculated
earlier in this framework \cite{eks5,egs98} (see also
\cite{eksann1,egs01r}).

\section{Diagrams with Radiative Photons in the Electron Line and
Electron Vacuum Polarization}

\begin{figure}[ht]
\centerline{\epsfig{file=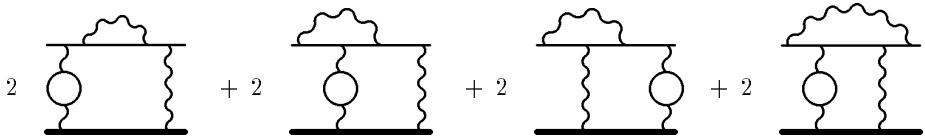,height=2cm}}
\vspace{0.5cm}
\caption{Electron line and electron vacuum polarization}
\label{ee}
\end{figure}

The general expression for the contribution to HFS arising from the
diagrams in Fig.\ \ref{ee}\footnote{The graphs with the crossed
exchanged photons are not shown explicitly in this figure and similar
figures below.} is obtained from the integral in
\eq{generalprev} by insertion in the integrand of the doubled one-loop
electron polarization $(\alpha/\pi)k^2I_e(k)$

\begin{equation}
2\left(\frac{\alpha}{\pi}\right)k^2I_e(k)=2\left(\frac{\alpha}{\pi}\right)
k^2\int_0^1dv\frac{v^2(1-v^2/3)}{4+k^2(1-v^2)},
\end{equation}

\noindent
where the additional multiplicity factor 2 corresponds to the fact
that we can insert the vacuum polarization in each of the exchanged
photons.

Then all contributions to HFS generated by the
diagrams in Fig.\ \ref{ee} are given by the integral

\beq   \label{elnepol}
\delta E^{\,ee} ~~=~~  \frac{\alpha^2
(Z\alpha)}{\pi^3}\frac{m}{M} \widetilde E_F~ \frac{1}{4\pi \mu^2} ~
\int_0^{\pi}
{d \theta}~ \sin^2 {\theta} ~ \int_0^{\infty} {d k^2}~
~{\cal D}(k, \theta)
\eeq
\[
\frac{k^4}{k^2 + \mu^{-2} ~\cos^2 {\theta}}
~ \frac{1}{~(k^2 + a^2)^2 ~+~ 4 b^2 k^2 \cos^2{\theta}~}~
\int_0^1dv\frac{v^2(1-v^2/3)}{4+k^2(1-v^2)}.
\]

\noindent
To obtain the radiative-recoil corrections of order
$\alpha^2(Z\alpha)(m/M)\widetilde E_F$ contained in this integral we
follow the route described in the previous section.  First we write
the integral in \eq{elnepol} as a sum of seven $\mu$- and seven
$c$-integrals. These integrals are collected in Table \ref{table2} and
Table \ref{table3}, where the dimensionless contributions $\delta
\epsilon$ to the energy shifts are defined as $\delta E_i=\delta
\epsilon_i\,\alpha^2(Z\alpha)(m/M)\widetilde E_F/\pi^3$.

The $c$-integrals in Table \ref{table3} automatically contain only
contributions linear in the recoil factor $m/M$, and can be immediately
calculated numerically. The respective results are again presented in
Table \ref{table3}, and the total contribution of all $c$-integrals is

\beq   \label{totcellelpol}
\delta E^{ee}_{c}=\delta \epsilon^{ee}_{c}\,\frac{\alpha^2
(Z\alpha)}{\pi^3}\frac{m}{M} \widetilde E_F
=6.96182~(3)\frac{\alpha^2
(Z\alpha)}{\pi^3}\frac{m}{M} \widetilde E_F.
\eeq

The situation with the $\mu$-integrals is more complicated. Besides the
recoil contributions linear in the small mass ratio $m/M$, they contain
both nonrecoil contributions and the recoil contributions of higher
order in the mass ratio. We would like to remove the already known
nonrecoil contributions, to extract in analytic form all
coefficients before the logarithmically enhanced terms linear in the recoil
factor, and to calculate numerically the nonlogarthmic term
which is linear in the recoil factor. To this end, we need to throw away
the recoil contributions of higher orders which are also contained in
the $\mu$-integrals and which are too small from the phenomenological
point of view. If preserved, these higher order recoil contributions
result only in the loss of accuracy in the numerical integrations. It
is easy to see that the integrals with the function $\Phi_0^{s}(k)$
generate only already known nonrecoil contributions and can be safely
omitted for our present goals. We extract the  contributions linear (and
logarithmically enhanced) in the recoil factor from the
$\mu$-integrals by separating the integration region  with the help of
an auxiliary parameter $\sigma$ as described in the previous section.
All logarithms of the mass ratio originate from the high momentum parts
of the $\mu$-integrals, where we can use the high energy asymptotic
expansion of the polarization operator. Thus we obtain in the analytic
form all coefficients before the logarithms of the mass ratio. The
radiative-recoil corrections of order $\alpha(Z\alpha)(m/M)\widetilde
E_F$ in \eq{elnepol}, generated by the insertions of the radiative
photons in the electron line, are linear in the large logarithm of the
mass ratio. Hence the corrections of order
$\alpha^2(Z\alpha)(m/M)\widetilde E_F$ are quadratic in this large
logarithm. The coefficient before the logarithm squared, which we obtain
in this way, coincides with the one obtained earlier \cite{eks89}, and
the analytically obtained coefficient before the single-logarithmic
term, as well as the nonlogarithmic contribution, are new. The results
of the calculation of the $\mu$-integrals are collected in Tables
\ref{table4} and V.

As an illustration of our methods let us obtain the first three entries
in Table \ref{table4}. We start with the integral in the first line in
Table \ref{table2}. It contains a nonrecoil contribution

\beq
\delta \epsilon^{ee}_{\mu1}(non-recoil)=
\int_0^1 {dx} \int_0^x {dy}~ c_1 ~
\int_0^{\infty} {d k^2}~ \frac{k^4}{(k^2 + a^2)^3 }\Phi_0^{s}(k)
\int_0^1 {dv}\frac{v^2(1-v^2/3)}{4 + k^2(1-v^2)}
\eeq
\[
=\frac{1}{\mu}
\int_0^1 {dx} \int_0^x {dy}~ c_1 ~
\int_0^{\infty} {d k^2}~ \frac{k^3}{(k^2 + a^2)^3 }
\int_0^1 {dv}\frac{v^2(1-v^2/3)}{4 + k^2(1-v^2)}.
\]

\noindent
This integral, as well as all other nonrecoil contributions of order
$\alpha^2(Z\alpha) E_F$, was calculated analytically in \cite{eks1},
and numerically in \cite{kn1}, and we will not consider it here. We
write the recoil part of the integral in the first line in Table
\ref{table2} as a sum of low-momentum and high-momentum
integrals\footnote{Here, as well as in similar cases below, we allow
ourselves a slightly confusing notation, using the same symbol $\delta
\epsilon^{ee}_{\mu1}$ both for the total contribution of the
$\mu$-integral and for its recoil part.  We hope this will not lead to
any misunderstanding since in this paper we are interested only in the
recoil corrections.}

\beq
\delta \epsilon^{ee}_{\mu1}=\frac{1}{2}
\int_0^1 {dx} \int_0^x {dy} c_1
\int_0^{\infty} {d k^2} \frac{k^4}{(k^2 + a^2)^3 }
\Bigl[ 2\Phi_0^{\mu}(k)- \Phi_1^{\mu}(k) - \Phi_2^{\mu}(k)  \Bigr]
\int_0^1 {dv}\frac{v^2(1-v^2/3)}{4 + k^2(1-v^2)}
\eeq
\[
=\delta \epsilon^{ee>}_{\mu1}+\delta \epsilon^{ee<}_{\mu1}.
\]

\noindent
First we estimate the high momentum part $\delta \epsilon^{ee>}_{\mu1}$
considering the leading term in the integrand in the high momentum
limit (more technical details on similar estimates can be found in
\cite{egs98})

\beq
\delta \epsilon^{ee>}_{\mu1}=\frac{1}{2}
\int_0^1 {dx} \int_0^x {dy} c_1
\int_{\sigma^2}^{\infty} {d k^2} \frac{k^4}{(k^2 + a^2)^3 }
\Bigl[ 2\Phi_0^{\mu}(k)- \Phi_1^{\mu}(k) - \Phi_2^{\mu}(k)  \Bigr]
\int_0^1 {dv}\frac{v^2(1-v^2/3)}{4 + k^2(1-v^2)}
\eeq
\[
\sim ~~ \int_{\sigma}^{\infty} {\frac{d k^2}{k^4}} \ln^2{k^2}
~  ~ \Bigl[~ 2\Phi_0^{\mu}(k)~-~ \Phi_1^{\mu}(k)
~-~ \Phi_2^{\mu}(k) ~\Bigr]
~~ \sim ~~ \frac{1}{\sigma^2} \ln^2{\sigma}.
\]

\noindent
We see that the high momentum contribution is suppressed like
${1}/{\sigma^2}$. We have already extracted the recoil factor from
$\delta \epsilon^{ee}_{\mu1}$ explicitly, so now we are looking only
for such contributions which are not additionally suppressed.  Hence,
in the leading order in the recoil parameter we can omit the
high-momentum contrbution to $\delta \epsilon^{ee}_{\mu1}$, and in our
approximation the total contribution to the energy shift is given by
the low-momentum integral

\beq
\delta \epsilon^{ee<}_{\mu1}=\frac{1}{2}
\int_0^1 {dx} \int_0^x {dy} c_1
\int_0^{\sigma^2} {d k^2} \frac{k^4}{(k^2 + a^2)^3 }
\Bigl[ 2\Phi_0^{\mu}(k)- \Phi_1^{\mu}(k) - \Phi_2^{\mu}(k)  \Bigr]
\int_0^1 {dv}\frac{v^2(1-v^2/3)}{4 + k^2(1-v^2)}.
\eeq

\noindent
To extract the leading nonvanishing contribution to this integral at
$\mu\to0$ we substitute in the integrand the leading terms in the small
$\mu$ expansion of the functions $\Phi_i^{\mu}(k)$ (see Appendix
\ref{appa}) and obtain

\beq        \label{mu1inlow}
\delta \epsilon^{ee<}_{\mu1}\approx-\frac{21}{16}
\int_0^1 {dx} \int_0^x {dy} c_1
\int_0^{\sigma^2}{d k^2} \frac{k^4}{(k^2 + a^2)^3 }
\int_0^1 {dv}\frac{v^2(1-v^2/3)}{4 + k^2(1-v^2)}.
\eeq

\noindent
The momentum integral is ultravioletly convergent, so the magnitude
of the integral with an infinite upper limit differs from the integral
in \eq{mu1inlow} only by inverse powers of $\sigma$, which we omit
anyway. Hence, the total $\mu1$ contribution to the energy shift is
given by the integral

\beq
\delta \epsilon^{ee}_{\mu1}=-\frac{21}{16}
\int_0^1 {dx} \int_0^x {dy}~ c_1~
\int_{0}^{\infty} {d k^2} \frac{k^4}{(k^2 + a^2)^3 }
\int_0^1 {dv}\frac{v^2(1-v^2/3)}{4 + k^2(1-v^2)},
\eeq

\noindent
which is just the first entry in Table \ref{table4}.

Let us turn now to the next integrals in Table \ref{table2}. Each of
the integrals $\delta \epsilon^{ee}_{\mu2}$ and $\delta
\epsilon^{ee}_{\mu3}$ generates spurious logarithm cubed contributions
which cancel in the sum of these terms.  It is therefore convenient to
rearrange these terms in such way that spurious logarithm cubed terms
do not arise at all. To this end we write the coefficient functions
$c_2$ in the form

\beq
c_2 = \frac{4(1-2y)}{y(1-y)}
 x(-1+x+2\ln{x})+ \frac{4}{(1-y)^3}
\Big[-(1-x)(-1-2y/x)
\eeq
\[
+2(-4+4y/x) \ln{x}
+ x(-1+x+2\ln{x}) (4-5y+2y^2)\Big]\equiv c_{2a} + c_{2b}.
\]

\noindent
We have chosen $c_{2a}$ essentially as the singular part of the
function $c_2$ at $y\simeq 0$ and multiplied this singular part by the
factor $(1-2y)/(1-y)$ in order to simplify integration over $y$. Since
$(1-2y)dy/(y(1-y))=-da^2/a^2$ this makes integration over $y$
trivial. The spurious logarithm cubed term originates only from the
integral of the function $c_{2a}$. On the other hand, the integrals
over $x$ and $y$ with the function $c_{2b}$ remain finite even if we
omit $a^2$ in the denominator $(k^2+a^2)^3$, which leads to significant
simplifications in the high-momentum part of the integral $\delta
\epsilon^{ee}_{\mu2}$.

Following the same logic, we represent the coefficient functions $c_3$
also as a sum of two functions  $c_{3a}$ and $c_{3b}$

\beq
c_3 =\frac{1-2y}{y(1-y)} (1-6x-2x^2)
+ \frac{1}{(1-y)^2}
\Big[2(1-y)(1-6x-2x^2) + (1-y)(8+2x)
\eeq
\[
- (1/x)(26(1-x)+2x(1-x) - 6(1-x^2)y/x - 6x(1-y) + 16 \ln{x})\Big]
\equiv c_{3a} + c_{3b}.
\]

\noindent
Now the sum of the  recoil contributions to HFS generated by the
${\mu2}$ and ${\mu3}$ integrals in Table \ref{table2} may
be written in the form

\beq
\delta \epsilon^{ee}_{\mu2}+\delta \epsilon^{ee}_{\mu3}
=\delta \epsilon^{ee}_{\mu2a+\mu3a}+\delta \epsilon^{ee}_{\mu2b}+
\delta \epsilon^{ee}_{\mu3b}.
\eeq

\noindent
Let us consider calculation of the $\delta\epsilon^{ee}_{\mu2a+\mu3a}$
contribution

\beq
\delta
\epsilon^{ee}_{\mu2a+\mu3a}=\frac{1}{2} \int_0^1 {dx} \int_0^x {dy}
\int_{0}^{\infty} {d k^2}   \Biggl[
\frac{c_{2a}k^6}{(k^2 + a^2)^3 }
- \frac{c_{3a}k^4}{(k^2 + a^2)^2 } \Biggr]
\eeq
\[
\int_0^1 {dv}
\frac{v^2(1-v^2/3)}{4 + k^2(1-v^2)}
\Bigl[ 2\Phi_0^{\mu}(k) +  \Phi_1^{\mu}(k) \Bigr]
=\epsilon^{ee>}_{\mu2a+\mu3a}+\epsilon^{ee<}_{\mu2a+\mu3a}.
\]

\noindent
As mentioned above, calculation of the $y$ integral becomes
trivial after the change of variables $y\to a^2$. In the high-momentum
integral we next expand over the inverse powers of $k^2$, and use the
asymptotic expansions of the arising functions $V_{lnm}$ (these
functions were defined and their properties were discussed in \cite{egs98})

\beq
\epsilon^{ee>}_{\mu2a+\mu3a}=\frac{1}{2}
\int_{\sigma^2}^{\infty} {\frac{d k^2}{k^2}}  \Biggl[
\biggl(-\frac{16}{3}\ln{k} - \frac{2\pi^2}{3} + 10 \biggr)
- \biggl(-\frac{16}{3}\ln{k} + \frac{20}{3} \biggr)
\Biggr]\biggl(\frac{2}{3}\ln{k} - \frac{5}{9} \biggr)
\eeq
\[
\Bigl[ 2\Phi_0^{\mu}(k) +  \Phi_1^{\mu}(k) \Bigr]
\approx  \biggl(-\frac{\pi^2}{6} + \frac{5}{6}  \biggr)
\biggl\{\frac{4}{3}  \biggl[2V_{110} + V_{111} \biggr] -
\frac{10}{9}  \biggl[2V_{100} + V_{101} \biggr]\biggr\}
\]
\[
\approx
\biggl(\frac{\pi^2}{3} - \frac{5}{3} \biggr)  \ln^2{\frac{M}{m}}
+ \biggl(\frac{4\pi^2}{9} -  \frac{20}{9} \biggr)  \ln{\frac{M}{m}}
+ \frac{\pi^4}{18} - \frac{5\pi^2}{18}
\]
\[
+ \biggl(-\frac{\pi^2}{3} + \frac{5}{3} \biggr)  \ln^2{\sigma}
+ \biggl(\frac{5\pi^2}{9} -  \frac{25}{9} \biggr)  \ln{\sigma}.
\]

\noindent
In the low-momentum part of the integral we again preserve only the
leading terms in the small $\mu$ expansion of the functions
$\Phi_i^{\mu}(k)$

\beq
\epsilon^{ee<}_{\mu2a+\mu3a}=\frac{3}{4} \int_0^1 {dx} \int_0^x {dy}
\int_{0}^{\sigma^2} {d k^2}  \Biggl[
-\frac{c_{2a}k^6}{(k^2 + a^2)^3 }
+ \frac{c_{3a}k^4}{(k^2 + a^2)^2 } \Biggr]
\int_0^1 {dv}
\frac{v^2(1-v^2/3)}{4 + k^2(1-v^2)}
\eeq

\noindent
and integrate explicitly over $y$

\beq
\epsilon^{ee<}_{\mu2a+\mu3a}=
\biggl(\frac{\pi^2}{3} - \frac{5}{3} \biggr)  \ln^2{\sigma}
+ \biggl(-\frac{5\pi^2}{9} +  \frac{25}{9} \biggr)  \ln{\sigma}
+\frac{14(\pi^2-5)}{27}
\eeq
\[
+ \frac{3}{4} \int_0^1 {dx}
\int_{0}^{\infty} {d k^2}
\Biggl\{(-1+6x+2x^2)
\biggl[\ln{k^2} - \ln{(k^2+a^2_1)} - \frac{a^2_1}{k^2+a^2_1}\biggr]
\]
\[
- 4x(-1+x+2\ln{x})
\biggl[-\ln{k^2} + \ln{(k^2+a^2_1)} + \frac{2a^2_1}{k^2+a^2_1}
- \frac{1}{2}  \frac{a^4_1}{(k^2+a^2_1)^2} \biggr]\Biggr\}
\]
\[
\int_0^1 {dv}
\frac{v^2(1-v^2/3)}{4 + k^2(1-v^2)}  \Biggr\},
\]

\noindent
where $a^2_1={x}/({1-x})$. Note that we again extended the momentum
integration to infinity in the finite integrals for the low-momentum
contribution. Finally, summing the high- and low-momentum contributions
we obtain the $\sigma$-independent result

\beq
\epsilon^{ee}_{\mu2a+\mu3a}=
\biggl(\frac{\pi^2}{3} - \frac{5}{3} \biggr)  \ln^2{\frac{M}{m}}
+ \biggl(\frac{4\pi^2}{9} -  \frac{20}{9} \biggr)  \ln{\frac{M}{m}}
+ \frac{\pi^4}{18} - \frac{5\pi^2}{18}+ \frac{14(\pi^2-5)}{27}
\eeq
\[
+ \frac{3}{4} \int_0^1 {dx}
\int_{0}^{\infty} {d k^2}
\Biggl\{(-1+6x+2x^2)
\biggl[\ln{k^2} - \ln{(k^2+a^2_1)} - \frac{a^2_1}{k^2+a^2_1}\biggr]
\]
\[
- 4x(-1+x+2\ln{x})
\biggl[-\ln{k^2} + \ln{(k^2+a^2_1)} + \frac{2a^2_1}{k^2+a^2_1}
- \frac{1}{2}  \frac{a^4_1}{(k^2+a^2_1)^2} \biggr]\Biggr\}
\]
\[
\int_0^1 {dv}
\frac{v^2(1-v^2/3)}{4 + k^2(1-v^2)}  \Biggr\},
\]

\noindent
which is convenient for numerical calculations.

As our final example let us derive an expression
for $\delta\epsilon^{ee}_{\mu2b}$ in the third line in Table
\ref{table4}. We start with the general expression for this
contribution from Table \ref{table2}

\beq  \label{mu2bgenr}
\delta
\epsilon^{ee}_{\mu2b}=\frac{1}{2} \int_0^1 {dx} \int_0^x {dy}
\int_{0}^{\infty} {d k^2} \frac{c_{2b}k^6}{(k^2 + a^2)^3 }
\int_0^1 {dv}
\frac{v^2(1-v^2/3)}{4 + k^2(1-v^2)}
\Bigl[ 2\Phi_0^{\mu}(k) +  \Phi_1^{\mu}(k) \Bigr].
\eeq

\noindent
As we already mentioned above in this case we are free to  omit $a^2$ in
comparison with $k^2$ in the denominator in the high-momentum
part of this integral

\beq
\epsilon^{ee>}_{\mu2b}=\frac{1}{2} \int_0^1 {dx} \int_0^x {dy}
\int_{\sigma^2}^{\infty} {d k^2}
{c_{2b}}
 \int_0^1 {dv}
\frac{v^2(1-v^2/3)}{4 + k^2(1-v^2)}
\Bigl[ 2\Phi_0^{\mu}(k) +  \Phi_1^{\mu}(k) \Bigr]
\eeq
\[
\approx  \frac{1}{2}
\int_{\sigma}^{\infty} {\frac{d k^2}{k^2}}
\biggl( \frac{10\pi^2}{3} - \frac{271}{9}  \biggr)
   \biggl(\frac{2}{3}\ln{k} - \frac{5}{9} \biggr)
\Bigl[ 2\Phi_0^{\mu}(k) +  \Phi_1^{\mu}(k) \Bigr]
\]
\[
\approx  \biggl(\frac{5\pi^2}{6} - \frac{271}{36}  \biggr)
\biggl\{\frac{4}{3}  \biggl[2V_{110} + V_{111} \biggr] -
\frac{10}{9}  \biggl[2V_{100} + V_{101} \biggr]\biggr\}
\]
\[
\approx
\biggl(-\frac{5\pi^2}{3} + \frac{271}{18} \biggr)  \ln^2{\frac{M}{m}}
+ \biggl(-\frac{20\pi^2}{9} + \frac{542}{27} \biggr)  \ln{\frac{M}{m}}
- \frac{5\pi^4}{18} + \frac{271\pi^2}{108}
\]
\[
+ \biggl(\frac{5\pi^2}{3} - \frac{271}{18} \biggr)  \ln^2{\sigma}
+ \biggl(-\frac{25\pi^2}{9} +  \frac{1355}{54} \biggr)
\ln{\sigma}.
\]

Dealing with the low-momentum part of the integral in \eq{mu2bgenr} we
would like to extract analytically leading logarithms of $\sigma$. This
would allow us to get rid of any trace of the parameter $\sigma$ in the
final expression for this integral, making it much more suitable for
further numerical calculations. The double integral over $x$
and $y$, with the integrand including the function $c_{2b}$, cannot be
calculated analytically as easily as the integral with the function
$\epsilon^{ee}_{\mu2a+\mu3a}$ above. In order to overcome this
difficulty we use the identity

\beq  \label{siplxy}
\frac{k^4}{(k^2 + a^2)^3} = \frac{1}{k^2 + 4} +
\biggl[\frac{k^4}{(k^2 + a^2)^3} - \frac{1}{k^2 + 4}\biggr]
\eeq

\noindent
in the low-momentum part of the integral in \eq{mu2bgenr}.
Analytic calculation of the integral with the first term  on the right
hand side (RHS) is simple. On the other hand, all logarithms of $\sigma$
are supplied by this integral since the second term on the RHS in
\eq{siplxy} decreases at large $k$ faster than the first term. Then we
obtain the low-momentum contribution to the integral in
\eq{mu2bgenr} in the form

\beq
\delta
\epsilon^{ee<}_{\mu2b}=\biggl(-\frac{5\pi^2}{3} + \frac{271}{18}
\biggr)  \ln^2{\sigma} + \biggl(\frac{25\pi^2}{9} -  \frac{1355}{54}
\biggr) \ln{\sigma}
-\frac{1}{54} \bigl( 30\pi^2-271 \bigr)
 \bigl( 8\ln{2}-3\bigr)
\eeq
\[
- \frac{3}{4} \int_0^1 {dx} \int_0^x {dy} c_{2b}
\int_{0}^{\infty} {d k^2}
\biggl[\frac{k^6}{(k^2 + a^2)^3} - \frac{k^2}{k^2 + 4}\biggr]
  \int_0^1 {dv}
\frac{v^2(1-v^2/3)}{4 + k^2(1-v^2)}.
\]

\noindent
Finally, the total expression for the $\mu2b$ integral has the form

\beq
\delta \epsilon^{ee}_{\mu2b} =
\biggl(-\frac{5\pi^2}{3} + \frac{271}{18} \biggr)  \ln^2{\frac{M}{m}}
+ \biggl(-\frac{20\pi^2}{9} + \frac{542}{27} \biggr)  \ln{\frac{M}{m}}
\eeq
\[
- \frac{5\pi^4}{18} + \frac{271\pi^2}{108}
-\frac{1}{54} \bigl( 30\pi^2-271 \bigr)
\bigl( 8\ln{2}-3\bigr)
\]
\[
- \frac{3}{4} \int_0^1 {dx} \int_0^x {dy} c_{2b}
\int_{0}^{\infty} {d k^2}
\biggl[\frac{k^6}{(k^2 + a^2)^3} - \frac{k^2}{k^2 + 4}\biggr]
\int_0^1 {dv}
\frac{v^2(1-v^2/3)}{4 + k^2(1-v^2)}.
\]

Calculation of the remaining $\mu$-integrals goes along the similar
lines, and does not require any additional new tricks. Calculating the
nonlogarithmic contributions numerically and collecting all
$\mu$-integrals from Table \ref{table4}, we obtain the total
contribution of all $\mu$-integrals

\beq  \label{totmuellelpol}
\delta \epsilon^{ee}_\mu=
\frac{5}{2} \ln^2{\frac{M}{m}}
+ \frac{22}{3} \ln{\frac{M}{m}} + 4.~45606~(1).
\eeq

\noindent
The sum of $\mu$- and $c$-integrals in
\eq{totmuellelpol} and  \eq{totcellelpol} gives all radiative-recoil
corrections of order $\alpha^2(Z\alpha)(m/M)\widetilde E_F$ generated
by the diagrams in Fig.\ \ref{ee}

\beq \label{eeresult}
\delta\epsilon^{ee}=\delta \epsilon^{ee}_{\mu}+\delta \epsilon^{ee}_{c}=
\frac{5}{2} \ln^2{\frac{M}{m}}
+\frac{22}{3}\ln{\frac{M}{m}}+ 11.41788~(3).
\eeq

\section{Diagrams with Radiative Photons  in the Electron Line and
Muon Vacuum Polarization}

\begin{figure}[ht]
\centerline{\epsfig{file=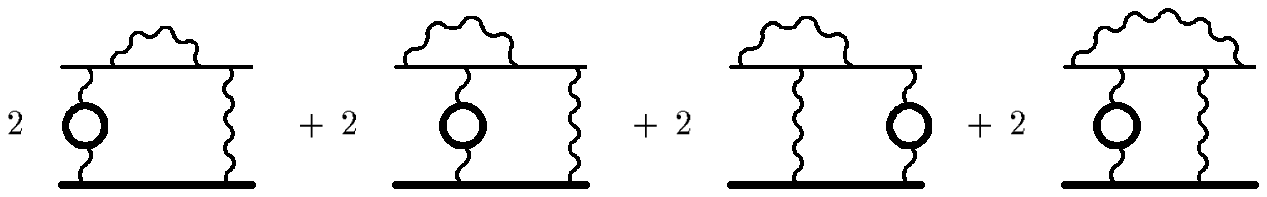,height=2cm}}
\vspace{0.5cm}
\caption{Electron line and muon vacuum polarization}
\label{me}
\end{figure}

Let us consider now the diagrams in Fig.\ \ref{me}. The only difference
between these diagrams and the diagrams in  Fig.\ \ref{ee} from  the
previous section is that they contain the muon vacuum polarization
insertion\footnote{We ascribe an extra factor $Z$ to each photon
emission by the heavy line, but being somewhat inconsequential, do not
write any $Z$ factor in the muon vacuum polarization.}

\begin{equation}         \label{muoninsertioin}
2\left(\frac{\alpha}{\pi}\right)k^2I_\mu(k)
=2\left(\frac{\alpha}{\pi}\right)k^2
\int_0^1dv\frac{v^2(1-v^2/3)}{\mu^{-2}+k^2(1-v^2)},
\end{equation}

\noindent
instead of the electron vacuum polarization, and the respective
contribution to HFS has the form (compare \eq{elnepol})

\beq   \label{elnemupol}
\delta E^{e\mu} ~~=~~  \frac{\alpha^2
(Z\alpha)}{\pi^3}\frac{m}{M} \widetilde E_F~ \frac{1}{4\pi \mu^2} ~
\int_0^{\pi}
{d \theta}~ \sin^2 {\theta} ~ \int_0^{\infty} {d k^2}~
~{\cal D}(k, \theta)
\eeq
\[
\frac{k^4 }{k^2 + \mu^{-2} ~\cos^2 {\theta}}
~ \frac{1}{~(k^2 + a^2)^2 ~+~ 4 b^2 k^2 \cos^2{\theta}~}~~
\int_0^1dv\frac{v^2(1-v^2/3)}{\mu^{-2}+k^2(1-v^2)}.
\]

\noindent
We transform the integrand in the sum of $\mu$- and $c$-integrals
exactly as in the previous section, the only difference being that each
integrand now contains as a factor the muon vacuum polarization
insertion from \eq{muoninsertioin} instead of the electron vacuum
polarization insertion. This immediately leads to significant
simplification of further calculations in comparison to the case of
the electron polarization. The muon polarization insertion in
\eq{elnemupol} is suppressed as $k^2\mu^2$ at integration momenta
much smaller than the muon mass. But the characteristic integration
momenta in the $c$-integrals are of order one in our dimensionless
units, and hence now all $c$-integralsdo not generate corrections
linear in the recoil factor $m/M$. Low-momenta parts of the
$\mu$-integrals, where integration goes over momenta
$k\leq\sigma\ll1/\mu$, are also suppressed as $\sigma^2\mu^2$. Only the
high-momenta parts of the $\mu$-integrals, where integration effectively
goes over the momenta comparable to $1/\mu$, generate contributions
linear in $m/M=2\mu$. Such contributions are present only in the
$\mu$-integrals connected with the coefficient functions $c_2$, $c_3$,
$c_6$, and $c_7$. Note  that the integrals with the coefficient
functions $c_2$ and $c_3$ contain the function  $\Phi_0^{s}(k)=1/(\mu
k)$ in the integrand. In the case of the electron vacuum polarization
(and absence of any polarization at all) the integrals with this
function generated nonrecoil corrections only. This essentially
happened because the characteristic integration momenta were of order
one, and in this region the contribution of the function  $1/(\mu k)$
is obviously enhanced as $1/\mu$. In the present case, characteristic
integration momenta are about $1/(\mu k)$, and the integrals with the
function $\Phi_0^{s}(k)$ generate recoil corrections on a par with the
integrals with the other functions $\Phi(k)$.

Integration over the Feynman parameters $x$ and $y$
in the high-momenta integrals simplifies and may be performed
analytically, and we obtain all one-dimensional momenta integrals which
generate radiative-recoil corrections in the case of muon polarization
insertions

\beq
\delta \epsilon^{e\mu>}_{\mu2}\approx
2\int_{\sigma^2}^{\infty} {d k^2} \biggl[
\biggl(-\frac{4}{3}\ln{k} - \frac{\pi^2}{6} + \frac{5}{2} \biggr)+
\biggl(\frac{5\pi^2}{6} - \frac{271}{36} \biggr) \bigg]
   \Bigl[2\Phi_0^{s}(k)+ 2\Phi_0^{\mu}(k) +  \Phi_1^{\mu}(k) \Bigr]
\eeq
\[
\int_0^1dv\frac{v^2(1-v^2/3)}{\mu^{-2}+k^2(1-v^2)},
\]
\beq
\delta \epsilon^{e\mu>}_{\mu3} \approx 2\int_{\sigma^2}^{\infty}
{d k^2} \biggl[
\biggl(\frac{4}{3}\ln{k} - \frac{5}{3} \biggr)+
\biggl(-\frac{2\pi^2}{3} + \frac{89}{18} \biggr) \bigg]
   \Bigl[2\Phi_0^{s}(k)+2\Phi_0^{\mu}(k) +  \Phi_1^{\mu}(k) \Bigr]
\eeq
\[
\int_0^1dv\frac{v^2(1-v^2/3)}{\mu^{-2}+k^2(1-v^2)},
\]
\beq
\delta \epsilon^{e\mu>}_{\mu6} \approx
\biggl({\pi^2} - \frac{21}{2} \biggr)
\int_{\sigma}^{\infty} {d k^2}
   \Phi_1^{\mu}(k)
\int_0^1dv\frac{v^2(1-v^2/3)}{\mu^{-2}+k^2(1-v^2)},
\eeq
\beq
\delta \epsilon^{e\mu>}_{\mu7}  \approx
\biggl(-{\pi^2} + \frac{15}{2} \biggr)
\int_{\sigma}^{\infty} {d k^2} \Phi_1^{\mu}(k)
\int_0^1dv\frac{v^2(1-v^2/3)}{\mu^{-2}+k^2(1-v^2)}.
\eeq

\noindent
The sum of all these contributions has the form

\beq
\delta \epsilon^{e\mu>}_\mu \approx
-\int_{\sigma^2}^{\infty} {d k^2} \biggl\{
\frac{7}{2}    \Bigl[ 2\Phi_0^s(k)
+ 2\Phi_0^{\mu}(k)
+  \Phi_1^{\mu}(k) \Bigr]
+ {3}    \Phi_1^{\mu}(k) \biggr\}
\int_0^1dv\frac{v^2(1-v^2/3)}{\mu^{-2}+k^2(1-v^2)},
\eeq

\noindent
or explicitly

\beq
\delta \epsilon^{e\mu>}_\mu \approx
-\int_{\sigma^2}^{\infty} {d k^2} \biggl\{
\frac{7}{2}    \biggl[
\frac{2}{\mu k}\biggl(\sqrt{1+\mu^2k^2} - \mu k  \biggr)
+ \biggl(-\mu k \sqrt{1+\mu^2k^2} + \mu^2 k^2
+ \frac{1}{2} \biggr)\biggr]
\eeq
\[
+{3}  \biggl(-\mu k \sqrt{1+\mu^2k^2} + \mu^2 k^2
+ \frac{1}{2} \biggr) \biggr\}
\int_0^1dv\frac{v^2(1-v^2/3)}{\mu^{-2}+k^2(1-v^2)}.
\]

\noindent
We can extend the momentum integration region in this integral to zero,
since contributions from small momenta are additionally suppressed by
powers of $\sigma\mu$. It is also natural to rescale the integration
momentum $k\to \mu k$. Then the expression for the  radiative-recoil
corrections has the form

\beq
\delta \epsilon^{e\mu}_\mu=-\int_{0}^{\infty} {d k} \biggl\{
{7}    \biggl[
{2}\biggl(\sqrt{1+k^2} -  k  \biggr)
+k \biggl(- k \sqrt{1+k^2} + k^2
+ \frac{1}{2} \biggr)\biggr]
\eeq
\[
+ {6} k \biggl(- k \sqrt{1+k^2} + k^2
+ \frac{1}{2} \biggr) \biggr\}
\int_0^1dv\frac{v^2(1-v^2/3)}{1+k^2(1-v^2)}.
\]

The integral in the first line of this equation is proportional to the
integral for the muon polarization contribution of order
$\alpha(Z\alpha)(m/M)\widetilde E_F$ in \cite{egs98}, and the
contribution of the integral in the second line may be calculated in
the same way. Finally we obtain

\beq   \label{emuresult}
\delta \epsilon^{e\mu}_\mu ~~=~~ -\frac{5\pi^2}{12} ~+~ \frac{1}{18}
~~=~~ -~4.~0567796\ldots ~~.
\eeq

\section{Diagrams with Radiative Photons  in the Muon Line and
Muon Vacuum Polarization}

\begin{figure}[ht]
\centerline{\epsfig{file=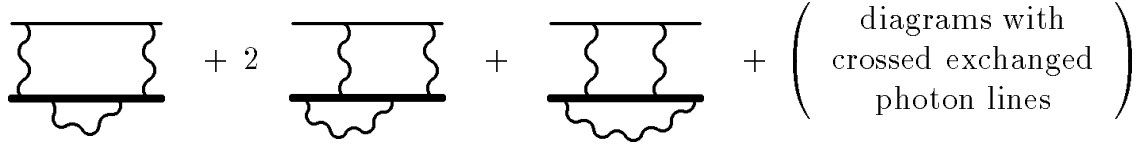,height=1.9cm}}
\vspace{0.5cm}
\caption{Muon-line radiative-recoil corrections}
\label{muonlineradrechfsfir}
\end{figure}

The  expression in \eq{electrcontr}, for the radiative corrections
generated by the diagrams with the radiative photon insertions in the
electron line Fig.\ \ref{ellineradreclamb}, was obtained without
expansion in the mass ratio.  Hence, after the substitutions
$m\leftrightarrow M$ and $\alpha\to Z^2\alpha$ it goes into the
expression for the corrections generated by the diagrams in Fig.\
\ref{muonlineradrechfsfir}\footnote{We use the convention that the
heavy line charge is $Ze$.}

\beq  \label{muonradphot}
\delta E^{\mu-line} ~~=~~
\frac{(Z^2\alpha)(Z\alpha)}{\pi^2} \frac{m}{M} ~ E_F ~  ~
\frac{1}{4} \int_0^1 {dx} \int_0^x {dy}
\int \frac{d^4 k}{i \pi^2(k^2 + i0)^2}~
\eeq
\[
\biggl(
\frac{1}{k^2 + 4\mu k_0 + i0}
~+~ \frac{1}{k^2 - 4\mu k_0 + i0}
\biggr)
\]

\[
 \biggl\{~ (~3k^2_0 ~-~ 2{\bf k}^2 ~)
\biggl[~\frac{c_{1}  {\bf k}^2 ~+~ c_{2}  (k^2)^2}
{(~-k^2 ~+~ 2bk_0 ~+~ a^2~)^3}
~+~\frac{ c_{3}  k^2 ~+~ c_{4}  2k_0}
{(~-k^2 ~+~ 2bk_0 ~+~ a^2~)^2} \biggr]
\]

\[
~-~ 3k_0  \biggl[
~\frac{c_{5}  k^2 ~+~ c_{6}  k^2  2k_0}
{(~-k^2 ~+~ 2bk_0 ~+~ a^2~)^2}
~+~ \frac{c_{7}  k^2}
{~-k^2 ~+~ 2bk_0 ~+~ a^2~} \biggr] \biggr\} .
\]

\noindent
The dimensionless integration momentum here is measured in muon
mass units, and apparent incomplete symmetry with the expression in
\eq{electrcontr} is due to asymmetry of the parameter $\mu=m/(2M)$.

Unlike the case of the radiative photon insertions in the electron line
in Section \ref{radinsel}, the integral in \eq{muonradphot} does not
generate nonrecoil contributions \cite{eks6,eksann2}. This happens because
the radiative photon insertions in the muon line suppress the
contribution from the integration momenta of order of the electron mass
which were responsible for the nonrecoil correction by an extra factor
$\mu^2$.  Hence, for calculation of the leading order recoil
corrections we can safely allow $\mu\to0$ in \eq{muonradphot}

\beq  \label{muonradphotmu0}
\delta E^{\mu-line} ~~\approx~~
\frac{(Z^2\alpha)(Z\alpha)}{\pi^2} \frac{m}{M} ~ E_F ~  ~
\frac{1}{2} \int_0^1 {dx} \int_0^x {dy}
\int \frac{d^4 k}{i \pi^2(k^2 + i0)^3}~
\eeq
\[
 \biggl\{~ (~3k^2_0 ~-~ 2{\bf k}^2 ~)
\biggl[~\frac{c_{1}  {\bf k}^2 ~+~ c_{2}  (k^2)^2}
{(~-k^2 ~+~ 2bk_0 ~+~ a^2~)^3}
~+~\frac{ c_{3}  k^2 ~+~ c_{4}  2k_0}
{(~-k^2 ~+~ 2bk_0 ~+~ a^2~)^2} \biggr]
\]

\[
~-~ 3k_0  \biggl[
~\frac{c_{5}  k^2 ~+~ c_{6}  k^2  2k_0}
{(~-k^2 ~+~ 2bk_0 ~+~ a^2~)^2}
~+~ \frac{c_{7}  k^2}
{~-k^2 ~+~ 2bk_0 ~+~ a^2~} \biggr] \biggr\} .
\]

\noindent
After the Wick rotation and transition to the four-dimensional
spherical coordinates we have

\beq   \label{muonradphotmueuc}
\delta E^{\mu-line} =
\frac{(Z^2\alpha)(Z\alpha)}{\pi^2} \frac{m}{M}  E_F
\frac{1}{2\pi}
\int_0^{\pi} {d \theta} \sin^2 {\theta}
 \int_0^{\infty} {d k^2}{\cal D}(k, \theta)
 \frac{1}{(k^2 + a^2)^2 + 4 b^2 k^2 \cos^2{\theta}},
\eeq
\[
\]

\noindent
where the differential operator ${\cal D}(k,\theta)$ was defined in
\eq{difop}. There are no $\mu$-dependent terms in
\eq{muonradphotmueuc}, and this expression for the energy shift
is similar to the $c$-integrals in \eq{generalprevsep}. It can
be formally obtained from the expression in \eq{generalprevsep} by the
substitution

\[
-~ \frac{b^2 k^4}{(k^2 + a^2)^2} ~~ \to  Z^2 ~~.
\]

\begin{figure}[ht]
\centerline{\epsfig{file=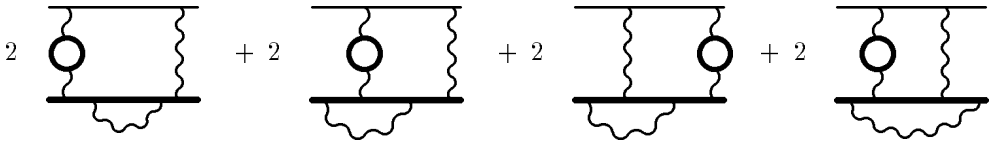,height=2cm}}
\vspace{0.5cm}
\caption{Muon line and muon vacuum polarization}
\label{mm}
\end{figure}

All corrections to HFS described by the integral in
\eq{muonradphotmueuc} were analytically calculated in
\cite{eks6}. Contributions to the energy shift generated by the
diagrams in Fig.\ \ref{mm} are obtained from the expression in
\eq{muonradphotmueuc} by insertion in the integrand of the doubled muon
vacuum polarization

\begin{equation}         \label{muoninsertioinmu}
2\left(\frac{\alpha}{\pi}\right)k^2I_\mu(k)
=2\left(\frac{\alpha}{\pi}\right)k^2
\int_0^1dv\frac{v^2(1-v^2/3)}{4+k^2(1-v^2)}.
\end{equation}

\noindent
This expression  is apparently different from the respective
expression in \eq{muoninsertioin} because now our dimensionless momenta
are measured in terms of the muon mass. Finally, the
integral for the contribution to HFS generated by the diagrams in Fig.\
\ref{mm} has the form

\beq   \label{muonradphotmueucmupol}
\delta E^{\mu\mu} ~~=~~
\frac{\alpha(Z^2\alpha)(Z\alpha)}{\pi^3} \frac{m}{M} ~ E_F ~  ~
\frac{1}{\pi}
\int_0^{\pi} {d \theta}~ \sin^2 {\theta}
~ \int_0^{\infty} {d k^2}~{\cal D}(k, \theta)
\eeq
\[
~ \frac{k^2}{~(k^2 + a^2)^2 ~+~ 4 b^2 k^2 \cos^2{\theta}~}
\int_0^1dv\frac{v^2(1-v^2/3)}{4+k^2(1-v^2)}
=\left(\sum_1^7\delta \epsilon^{\mu \mu}_i\right)
\frac{\alpha(Z^2\alpha)(Z\alpha)}{\pi^3} \frac{m}{M} ~ E_F.
\]

\noindent
All seven contributions are collected in Table \ref{table5}, and the
total contribution to HFS generated by the diagrams in Fig.\ \ref{mm} is

\beq  \label{mumuresult}
\delta \epsilon^{\mu\mu}=-1.~80176~(2).
\eeq

\section{Diagrams with Radiative Photons in the Muon Line and
Electron Vacuum Polarization}

\begin{figure}[ht]
\centerline{\epsfig{file=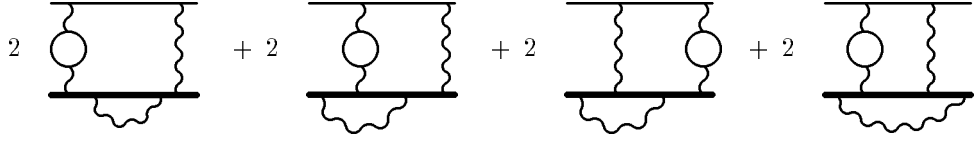,height=2cm}}
\vspace{0.5cm}
\caption{Muon line and electron vacuum polarization}
\label{em}
\end{figure}

Let us turn now to the diagrams in Fig.\ \ref{em}. The only difference
between these diagrams and the diagrams in Fig.\ \ref{mm} is that now we
have electron polarization insertions instead of the muon polarization
insertions. The respective analytic expression is obtained from the one
in \eq{muonradphotmueucmupol} by substitution of the electron
polarization instead of the muon polarization

\beq
2 k^2 \int_0^1 {dv}~
\frac{v^2(1-v^2/3)}{4 + k^2(1-v^2)} ~~\to ~~
2 k^2 \int_0^1 {dv}~
\frac{v^2(1-v^2/3)}{4(m/M)^2 + k^2(1-v^2)}.
\eeq

\noindent
The typical integration momenta in \eq{muonradphotmueuc} are of order
of the muon mass and, since we are calculating to linear accuracy in
$m/M$, it is sufficient to substitute instead of the electron
polarization operator the leading terms in its expansion over $m/M$

\beq
2 k^2 \int_0^1 {dv}~
\frac{v^2(1-v^2/3)}{4(m/M)^2 + k^2(1-v^2)} ~~\approx ~~~
\frac{2}{3}\ln{k^2}+\frac{4}{3} \ln{\frac{M}{m}}
 ~-~ \frac{10}{9} ~~.
\eeq

\noindent
Then the total contribution to HFS generated by the diagrams in Fig.\
\ref{em} is given by the integral

\beq   \label{muonradphotmueucepol}
\delta E^{\mu e} ~~=~~
\frac{\alpha(Z^2\alpha)(Z\alpha)}{\pi^3} \frac{m}{M} ~ E_F ~  ~
\frac{1}{2\pi}
\int_0^{\pi} {d \theta}~ \sin^2 {\theta}
~ \int_0^{\infty} {d k^2}~{\cal D}(k, \theta)
\eeq
\[
~ \frac{1}{~(k^2 + a^2)^2 ~+~ 4 b^2 k^2 \cos^2{\theta}~}~~
\left(\frac{2}{3}\ln{k^2}+\frac{4}{3} \ln{\frac{M}{m}}
 ~-~ \frac{10}{9}\right)
\]
\[
=\left(\delta \epsilon^{\mu e}_c+\delta \epsilon^{\mu e}_s\right)
\frac{\alpha(Z^2\alpha)(Z\alpha)}{\pi^3} \frac{m}{M} ~ E_F,
\]

\noindent
where the contribution $\delta \epsilon^{\mu e}_c=\sum_i\epsilon^{\mu
e}_{ci}$ corresponds to the first term in the last brackets, and
the contribution $\delta \epsilon^{\mu e}_s$ corresponds to the second
and third terms in the same brackets.  We calculate  $\delta
\epsilon^{\mu e}_c$ numerically and collect all seven contributions to
this integral in Table \ref{table6}. The sum of all these contributions
is

\beq
\delta \epsilon^{\mu e}_c =12.~94447~(4).
\eeq

The integral for $\delta \epsilon^{\mu e}_s$ is proportional to the
integral for the radiative-recoil corrections generated by the
radiative photon insertions in the muon line, which is known
analytically \cite{eks6,eksann2}. Hence, we can immediately put down an
analytic result for $\delta \epsilon^{\mu e}_s$

\beq
\delta \epsilon^{\mu e}_s=\biggl( \frac{4}{3}\ln{\frac{M}{m}}~~
-~~ \frac{10}{9}\biggr)
\biggl(~\frac{9}{2}~\zeta{(3)} ~-~ 3\pi^2 \ln{2} ~+~ \frac{39}{8}
~~\biggr).
\eeq

Then the total contribution to HFS generated by the diagrams in Fig.\
\ref{em} is equal to

\beq  \label{mueresult}
\delta  \epsilon^{\mu e} ~~~=~~~
\biggl(~6~\zeta{(3)} ~-~ 4 \pi^2 \ln{2} ~+~ \frac{13}{2}
~~\biggr) ~\ln{\frac{M}{m}} ~~+~~ 24.32115~(4).
\eeq

\section{Discussion of Results}

Collecting the results in \eq{eeresult}, \eq{emuresult},
\eq{mumuresult}, and \eq{mueresult} we obtain the
three-loop single--logarithmic and nonlogarithmic corrections
generated by the one-loop electron and muon polarization insertions in
the exchanged photons

\beq
\delta E=\delta \epsilon^{e}~\frac{\alpha^2
(Z\alpha)}{\pi^3}\frac{m}{M} \widetilde E_F+\delta  \epsilon^{\mu}~
\frac{\alpha(Z^2\alpha)
(Z\alpha)}{\pi^3}\frac{m}{M} \widetilde E_F,
\eeq

\noindent
where

\beq
\delta\epsilon^{e}=\frac{5}{2} \ln^2{\frac{M}{m}}
+\frac{22}{3}\ln{\frac{M}{m}}+7.36110~(3),
\eeq
\beq
\delta  \epsilon^{\mu}=\biggl(~6~\zeta{(3)} ~-~ 4 \pi^2 \ln{2} ~
+~ \frac{13}{2}
~~\biggr) ~\ln{\frac{M}{m}} ~~+~~22.51939~(5).
\eeq

\noindent
If we recall that in the real muonium $Z=1$ and that the logarithm
squared term was already calculated earlier \cite{eks89} then the new
three-loop single-logarithmic and nonlogarithmic corrections
obtained above may be written as

\beq
\delta E_{f}=\left[\biggl(~6~\zeta{(3)} ~-~ 4 \pi^2
\ln{2} ~ +~ \frac{83}{6}~~\biggr) ~\ln{\frac{M}{m}}
~~+~~29.88049~(6)\right] \frac{\alpha^3}{\pi^3}\frac{m}{M}
\widetilde E_F.
\eeq

\noindent
Combining this result with the result of our earlier paper
\cite{egs02} we obtain  all three-loop single--logarithmic and
nonlogarithmic corrections generated by the electron and muon
polarization insertions in the exchanged photons

\beq
\delta E_{tot}=\left[\biggl(~-~ 4 \pi^2
\ln{2} ~ +~ \frac{67}{12}~~\biggr) ~\ln{\frac{M}{m}}
~~+~~9.59318~(6)\right] \frac{\alpha^3}{\pi^3}\frac{m}{M}
\widetilde E_F,
\eeq

\noindent
or, numerically

\beq
\delta E=-0.~028~8\mbox{kHz}.
\eeq

This correction has the same scale as some other corrections to
hyperfine splitting in muonium calculated recently \cite{egs98,my,rh}.
All these results improve the accuracy of the theory of hyperfine
splitting and affect the value of the electron-muon mass ratio derived
from the experimental data \cite{lbdd} on hyperfine splitting (see,
e.g., reviews in \cite{egs01r,mt00}). We postpone discussion of the
phenomenological implications of the result above until the completion
of the calculations of the remaining three-loop radiative recoil
corrections.

\acknowledgements

This work was supported in part by the NSF grant PHY-0138210. Work of
V.A. Shelyuto was also supported by the Russian Foundation for Basic
Research.

\appendix
\section{Standard auxiliary functions}
\label{appa}

Calculations described in this paper are greatly facilitated by the
use of the auxiliary functions $~\Phi_n (k) ~$
($n=0,1,2,3$) defined by the relationship

\begin{equation}
\Phi_n (k) ~ \equiv ~  \frac{1}{\pi \mu^2}
\int_0^{\pi} {d \theta}~ \sin^2 {\theta}   \cos^{2n} {\theta}
~
\frac{~(k^2 + a^2)^2 ~-~ 4~ \mu^2~ b^2~ k^4~}
{\Bigr(~k^2 + \mu^{-2} ~\cos^2 {\theta} ~ \Bigl)
\Bigl[~(k^2 + a^2)^2 ~+~ 4~ b^2~ k^2 ~\cos^2{\theta}~\Bigr]}.
\end{equation}

\noindent
The integral over angles may be explicitly calculated, and the result of the
integration is conveniently written as a sum

\begin{equation}
\Phi_n (k) ~ \equiv ~ \Phi_n^S (k) ~+~\Phi_n^{\mu} (k) ~+~\Phi_n^C (k),
\end{equation}

\noindent
where

\begin{equation}
\Phi_n^S (k)  ~=~ \frac{\delta_{n0}}{\mu k} ,
\end{equation}
\begin{equation}
\Phi_0^{\mu} (k)  ~=~ W(\xi_{\mu}) ~-~ \frac{1}{\sqrt{\xi_{\mu}}},
\end{equation}
\begin{equation}
\Phi_1^{\mu} (k)  ~=~ -~\xi_{\mu}  W(\xi_{\mu}) ~+~ \frac{1}{2},
\end{equation}
\begin{equation}
\Phi_2^{\mu} (k)  ~=~ \xi_{\mu}  \Bigl(~\xi_{\mu}  W(\xi_{\mu})
~-~ \frac{1}{2} \Bigr) ~+~ \frac{1}{8} ,
\end{equation}
\begin{equation}
\Phi_3^{\mu} (k)  ~=~ -~ \xi_{\mu}  \Bigl[~\xi_{\mu}
\Bigl(~\xi_{\mu}  W(\xi_{\mu}) ~-~ \frac{1}{2} \Bigr) ~+~
\frac{1}{8}~\Bigr] ~+~ \frac {1}{16} ,
\end{equation}
\begin{equation}
\Phi_0^C (k)  ~=~-~ W(\xi_C) ,
\end{equation}
\begin{equation}
\Phi_1^C (k)  ~=~ \xi_C  W(\xi_C) ~-~ \frac{1}{2},
\end{equation}
\begin{equation}
\Phi_2^C (k)  ~=~-~\xi_C  \Bigl(~\xi_C  W(\xi_C)
~-~ \frac{1}{2} \Bigr) ~-~ \frac{1}{8} ,
\end{equation}
\begin{equation}
\Phi_3^C (k)  ~=~ \xi_C  \Bigl[~\xi_C
\Bigl(~\xi_C  W(\xi_C) ~-~ \frac{1}{2} \Bigr) ~+~
\frac{1}{8}~\Bigr] ~-~ \frac{1}{16} .
\end{equation}

\noindent
The standard function $W(\xi)$ has the form

\begin{equation}
W(\xi) ~=~ \sqrt {1 ~+~ \frac{1}{\xi}} ~-~ 1 .
\end{equation}

\noindent
and

\begin{equation}
\xi_{\mu} ~=~ \mu^2 ~ k^2 ,~~~~~
\xi_C ~=~ \frac{(k^2 + a^2)^2}{4 b^2 k^2}.
\end{equation}

One may easily obtain asymptotic expressions for the function $W(\xi)$

\begin{equation}
\lim_{\xi\rightarrow 0}W(\xi)\rightarrow\frac{1}{\sqrt{\xi}},
\end{equation}
\[
\lim_{ \xi\rightarrow \infty}W(\xi)\rightarrow\frac{1}{2\xi}.
\]

\noindent
High- and low-momentum asymptotic expressions for the functions
$\Phi_i(k)$ may also be easily calculated.  Let us cite low-momentum
expansions, which were used in the main text for calculation of the
contributions to the hyperfine splitting of relative order $\mu^2$

\begin{equation}
\Phi_0^{\mu} (k) ~\approx ~  -~1 ~+~ \frac{\mu k}{2} ,
\end{equation}
\begin{equation}
\Phi_1^{\mu} (k) ~\approx ~  \frac{1}{2} ~ - ~  \mu k  ~ + ~
(\mu k )^2,
\end{equation}
\begin{equation}
\Phi_2^{\mu} (k) ~\approx ~  \frac{1}{8} ~ - ~ \frac{(\mu k)^2}{2} .
\end{equation}

\begin{table}
\caption{Coefficients in the Electron-Line Factor}
\begin{tabular}{ll}
\\
$\mbox{c}_{1}$& $\frac{16}{y(1-y)^3}
\Big[~(1-x)(x-3y)~-~2y\ln{x}~\Big]~$
\\ \\ \tableline
\\
$\mbox{c}_{2}$      &   $\frac{4}{y(1-y)^3}
\Big[~-(1-x)(x - y - 2y^2 /x)~+~2(x - 4y + 4y^2 /x)  \ln{x}~\Big]~$
\\ \\ \tableline
\\
$\mbox{c}_{3}$     &   $\frac{1}{y(1-y)^2}
\Big[~1 - 6x - 2x^2 -(y/x) (26 - 6y/x - 37x - 2x^2
+ 12xy + 16 \ln{x})~\Big]~$
\\ \\ \tableline
\\
$\mbox{c}_{4}$   & $\frac{1}{y(1-y)^2}
\Big(~2x - 4x^2 - 5y + 7xy~ \Big)~$
\\ \\ \tableline
\\
$\mbox{c}_{5}$   & $\frac{1}{y(1-y)^2}
\Big(~ 6x - 3x^2 - 8y + 2xy ~\Big)~$
\\ \\ \tableline
\\
$\mbox{c}_{6}$   & $-b^2  \frac{x - y}{x^2}~$
\\ \\ \tableline
\\
$\mbox{c}_{7}$  & $2  \frac{1-x}{x}~$
\\ \\
\end{tabular}
\label{table1}
\end{table}

\begin{table}
\caption{$\mu$-integrals for the diagrams in Fig.\ \ref{ee}}
\begin{tabular}{ll}
\\
$\delta\epsilon^{ee}_{\mu1}$& $\frac{1}{2}
\int_0^1 {dx} \int_0^x {dy} c_1
\int_0^{\infty} {d k^2} \frac{k^4}{(k^2 + a^2)^3 }
\Bigl[ 2\Phi_0^{s}(k) + 2\Phi_0^{\mu}(k)
- \Phi_1^{\mu}(k) - \Phi_2^{\mu}(k)  \Bigr]
\int_0^1 {dv}\frac{v^2(1-v^2/3)}{4 + k^2(1-v^2)}$
\\ \\ \tableline
\\
$\delta\epsilon^{ee}_{\mu2}$      &  $\frac{1}{2}
\int_0^1 {dx} \int_0^x {dy} c_2
\int_0^{\infty} {d k^2} \frac{k^6}{(k^2 + a^2)^3 }
\Bigl[ 2\Phi_0^{s}(k) + 2\Phi_0^{\mu}(k)
+ \Phi_1^{\mu}(k)  \Bigr]
\int_0^1 {dv}\frac{v^2(1-v^2/3)}{4 + k^2(1-v^2)}$
\\ \\ \tableline
\\
$\delta\epsilon^{ee}_{\mu3}$     &  $-\frac{1}{2}
\int_0^1 {dx} \int_0^x {dy} c_3
\int_0^{\infty} {d k^2} \frac{k^4}{(k^2 + a^2)^2 }
\Bigl[ 2\Phi_0^{s}(k) + 2\Phi_0^{\mu}(k) +  \Phi_1^{\mu}(k)
\Bigr]\int_0^1 {dv}\frac{v^2(1-v^2/3)}{4 + k^2(1-v^2)}$
\\ \\ \tableline
\\
$\delta\epsilon^{ee}_{\mu4}$   &  $4\int_0^1 {dx} \int_0^x {dy} bc_4
\int_0^{\infty} {d k^2} \frac{k^4}{(k^2 + a^2)^3 }
\Bigl[ 2\Phi_1^{\mu}(k) + \Phi_2^{\mu}(k) \Bigr]
\int_0^1 {dv}\frac{v^2(1-v^2/3)}{4 + k^2(1-v^2)}$
\\ \\ \tableline
\\
$\delta\epsilon^{ee}_{\mu5}$  & $-6\int_0^1 {dx} \int_0^x {dy} bc_5
\int_0^{\infty} {d k^2} \frac{k^4}{(k^2 + a^2)^3 }
\Phi_1^{\mu}(k)\int_0^1 {dv}\frac{v^2(1-v^2/3)}{4 + k^2(1-v^2)}$
\\ \\ \tableline
\\
$\delta\epsilon^{ee}_{\mu6}$   & ${3}
\int_0^1 {dx} \int_0^x {dy} c_6
\int_0^{\infty} {d k^2} \frac{k^4}{(k^2 + a^2)^2 }
\Phi_1^{\mu}(k)\int_0^1 {dv}\frac{v^2(1-v^2/3)}{4 + k^2(1-v^2)}$
\\ \\ \tableline
\\
$\delta\epsilon^{ee}_{\mu7}$ & $-{3}
\int_0^1 {dx} \int_0^x {dy} bc_7
\int_0^{\infty} {d k^2} \frac{k^4}{(k^2 + a^2)^2 }
\Phi_1^{\mu}(k)\int_0^1 {dv}\frac{v^2(1-v^2/3)}{4 + k^2(1-v^2)}$
\\ \\
\end{tabular}
\label{table2}
\end{table}

\begin{table}
\caption{$c$-integrals for the diagrams in Fig.\ \ref{ee}}
\begin{tabular}{lll}
\\
$\delta\epsilon^{ee}_{c1}$& $-\frac{1}{4}
\int_0^1 {dx} \int_0^x {dy} c_1
\int_0^{\infty} {d k^2} \biggl(\frac{\partial}{\partial a^2}\biggr)^2
  \frac{k^4}{k^2 + a^2 }\frac{1}{\pi}\int_0^{\pi} {d \theta}
\sin^2 {\theta}$
\\  \\
& $(2 - \cos^{2} {\theta} - \cos^{4} {\theta})\frac{4 b^2 k^2}
{(k^2 + a^2)^2 + 4 b^2 k^2 \cos^2{\theta}}\int_0^1 {dv}
\frac{v^2(1-v^2/3)}{4 + k^2(1-v^2)}$ & -0.916881~(3)
\\ \\ \tableline
\\
$\delta\epsilon^{ee}_{c2}$      &   $- \frac{1}{4} \int_0^1 {dx}
\int_0^x {dy} c_2\int_0^{\infty} {d k^2}\biggl(\frac{\partial}{\partial
a^2}\biggr)^2\frac{k^6}{k^2 + a^2 }\frac{1}{\pi}
\int_0^{\pi} {d \theta}\sin^2 {\theta}
$
\\  \\
& $(2 + \cos^{2} {\theta})
\frac{4 b^2 k^2}{(k^2 + a^2)^2 ~+~ 4 b^2 k^2 \cos^2{\theta}}
\int_0^1 {dv}\frac{v^2(1-v^2/3)}{4 + k^2(1-v^2)}  $  &1.76474~(1)
\\ \\ \tableline
\\
$\delta\epsilon^{ee}_{c3}$     &   $- \frac{1}{2} \int_0^1 {dx}
\int_0^x {dy} c_3\int_0^{\infty} {d k^2} \frac{\partial}{\partial a^2}
\frac{k^4}{k^2 + a^2 }\frac{1}{\pi}
\int_0^{\pi} {d \theta} \sin^2 {\theta}  $
\\ \\
& $(2 + \cos^2{\theta})
\frac{4 b^2 k^2}{(k^2 + a^2)^2+4 b^2 k^2 \cos^2{\theta}}
\int_0^1 {dv}\frac{v^2(1-v^2/3)}{4 + k^2(1-v^2)}$ &5.93600~(5)
\\ \\ \tableline
\\
$\delta\epsilon^{ee}_{c4}$   & $2\int_0^1 {dx} \int_0^x {dy}~ bc_4
\int_0^{\infty} {d k^2}\frac{\partial}{\partial a^2}
\frac{k^4}{(k^2 + a^2)^2}\frac{1}{\pi}
\int_0^{\pi} {d \theta}\sin^2 {\theta} \cos^{2} {\theta}
$
\\ \\
& $(2+\cos^2{\theta})
\frac{4 b^2 k^2}{(k^2 + a^2)^2+4 b^2 k^2 \cos^2{\theta}}
\int_0^1 {dv}\frac{v^2(1-v^2/3)}{4 + k^2(1-v^2)} $ &0.0312952~(2)
\\ \\ \tableline
\\
$\delta\epsilon^{ee}_{c5}$   & $-{3}\int_0^1 {dx} \int_0^x {dy}bc_5
\int_0^{\infty} {d k^2}\frac{\partial}{\partial a^2}
\frac{k^4}{(k^2 + a^2)^2}\frac{1}{\pi}
\int_0^{\pi} {d \theta} \sin^2 {\theta}   $
\\ \\
& $\cos^{2} {\theta}
\frac{4 b^2 k^2}{(k^2 + a^2)^2 + 4 b^2 k^2 \cos^2{\theta}}
\int_0^1 {dv}\frac{v^2(1-v^2/3)}{4 + k^2(1-v^2)} $ &0.0003483~(2)
\\ \\ \tableline
\\
$\delta\epsilon^{ee}_{c6}$   & ${3}
\int_0^1 {dx} \int_0^x {dy}c_6\int_0^{\infty} {d k^2}~
\frac{\partial}{\partial a^2} ~
\frac{k^4}{k^2 + a^2 }\frac{1}{\pi}
\int_0^{\pi} {d \theta} \sin^2 {\theta}     $
\\ \\
& $\cos^{2} {\theta}
\frac{4 b^2 k^2}{(k^2 + a^2)^2 + 4 b^2 k^2 \cos^2{\theta}}
\int_0^1 {dv}\frac{v^2(1-v^2/3)}{4 + k^2(1-v^2)} $ & 0.0543901~(2)
\\ \\ \tableline
\\
$\delta\epsilon^{ee}_{c7}$ & ${3}\int_0^1 {dx} \int_0^x {dy}bc_7
\int_0^{\infty} {d k^2} \frac{k^4}{(k^2 + a^2)^2 }
\frac{1}{\pi}\int_0^{\pi} {d \theta}\sin^2 {\theta}    $
\\ \\
& $\cos^{2} {\theta}
\frac{4 b^2 k^2}{(k^2 + a^2)^2 + 4 b^2 k^2 \cos^2{\theta}}
\int_0^1 {dv}\frac{v^2(1-v^2/3)}{4 + k^2(1-v^2)}$ &0.0919342~(3)
\\ \\
\end{tabular}
\label{table3}
\end{table}

\begin{table}
\caption{Simplified $\mu$-integrals for the diagrams in Fig.\ \ref{ee}}
\begin{tabular}{lll}
\\
$\delta\epsilon^{ee}_{\mu1}$& $-\frac{21}{16}
\int_0^1 {dx} \int_0^x {dy} c_1
\int_0^{\infty} {d k^2} \frac{k^4}{(k^2 + a^2)^3 }
\int_0^1 {dv}\frac{v^2(1-v^2/3)}{4 + k^2(1-v^2)}
$& -2.105912~(2)
\\ \\ \tableline
\\
$\delta\epsilon^{ee}_{\mu2 a+\mu3 a }$      &   $\biggl(\frac{\pi^2}{3}
- \frac{5}{3}
\biggr)  \ln^2{\frac{M}{m}}
+ \biggl(\frac{4\pi^2}{9} -  \frac{20}{9} \biggr)  \ln{\frac{M}{m}}
+ \frac{\pi^4}{18} - \frac{5\pi^2}{18}+\frac{14(\pi^2-5)}{27}
$  & $\biggl(\frac{\pi^2}{3}
- \frac{5}{3}
\biggr)  \ln^2{\frac{M}{m}}$
\\  \\
& $
+ \frac{3}{4} \int_0^1 {dx}
\int_{0}^{\infty} {d k^2}
\Biggl\{(-1+6x+2x^2)
\biggl[\ln{k^2} - \ln{(k^2+a^2_1)}
$  & $+ \biggl(\frac{4\pi^2}{9} -  \frac{20}{9} \biggr)
\ln{\frac{M}{m}}$
\\  \\
& $
- \frac{a^2_1}{k^2+a^2_1}\biggr]- 4x(-1+x+2\ln{x})\biggl[-\ln{k^2}
+ \ln{(k^2+a^2_1)}
$
& $+ \frac{\pi^4}{18} - \frac{5\pi^2}{18}$
\\  \\
& $+ \frac{2a^2_1}{k^2+a^2_1}- \frac{1}{2}
 \frac{a^4_1}{(k^2+a^2_1)^2} \biggr]\Biggr\}
\int_0^1 {dv}\frac{v^2(1-v^2/3)}{4 + k^2(1-v^2)}
$  &+0.4698395~(7)
\\ \\ \tableline
\\
$\delta\epsilon^{ee}_{\mu2 b}$     &   $ \biggl(-\frac{5\pi^2}{3}
+ \frac{271}{18} \biggr)  \ln^2{\frac{M}{m}}
+ \biggl(-\frac{20\pi^2}{9} + \frac{542}{27} \biggr)  \ln{\frac{M}{m}}

$  & $\biggl(-\frac{5\pi^2}{3}
+ \frac{271}{18} \biggr)  \ln^2{\frac{M}{m}}$
\\ \\
& $
-\frac{1}{54} \bigl( 30\pi^2-271 \bigr)
 \bigl( 8\ln{2}-3\bigr)
- \frac{3}{4} \int_0^1 {dx} \int_0^x {dy} c_{2b}
$  &  $+ \biggl(-\frac{20\pi^2}{9} + \frac{542}{27} \biggr)
 \ln{\frac{M}{m}}$
\\  \\
& $
\int_{0}^{\infty} {d k^2}
\biggl[\frac{k^6}{(k^2 + a^2)^3} - \frac{k^2}{k^2 + 4}\biggr]
\int_0^1 {dv}\frac{v^2(1-v^2/3)}{4 + k^2(1-v^2)}
$ &$- \frac{5\pi^4}{18}+ \frac{271\pi^2}{108}$
\\  \\
&  &
-0.797177~(4)
\\ \\ \tableline
\\
$\delta\epsilon^{ee}_{\mu3 b}$   & $\biggl(\frac{4\pi^2}{3}
- \frac{89}{9} \biggr)  \ln^2{\frac{M}{m}}
+ \biggl(\frac{16\pi^2}{9} -  \frac{356}{27} \biggr)  \ln{\frac{M}{m}}
+ \frac{2\pi^4}{9} - \frac{89\pi^2}{54}
$ & $\biggl(\frac{4\pi^2}{3}
- \frac{89}{9} \biggr)  \ln^2{\frac{M}{m}}$
\\ \\
& $+
\frac{1}{27} \bigl( 12\pi^2-89\bigr)
 \bigl( 8\ln{2}-3\bigr)
+ \frac{3}{4} \int_0^1 {dx} \int_0^x {dy} c_{3b}
$  & $+ \biggl(\frac{16\pi^2}{9} -  \frac{356}{27} \biggr)
\ln{\frac{M}{m}}$
\\  \\
& $
\int_{0}^{\infty} {d k^2}
\biggl[\frac{k^4}{(k^2 + a^2)^2} - \frac{k^2}{k^2 + 4}\biggr]
\int_0^1 {dv}\frac{v^2(1-v^2/3)}{4 + k^2(1-v^2)} $ & $
+ \frac{2\pi^4}{9} - \frac{89\pi^2}{54}$
\\  \\
&   &
5.398568~(7)
\\ \\ \tableline
\\
$\delta\epsilon^{ee}_{\mu4 }$   & $\frac{9}{2}
\int_0^1 {dx} \int_0^x {dy} bc_4
\int_0^{\infty} {d k^2} \frac{k^4}{(k^2 + a^2)^3 }
\int_0^1 {dv}\frac{v^2(1-v^2/3)}{4 + k^2(1-v^2)}  $
& -0.3738824~(5)
\\ \\ \tableline
\\
$\delta\epsilon^{ee}_{\mu5}$   & $-{3}\int_0^1 {dx}
\int_0^x {dy} bc_5
\int_0^{\infty} {d k^2} \frac{k^4}{(k^2 + a^2)^3 }
\int_0^1 {dv}\frac{v^2(1-v^2/3)}{4 + k^2(1-v^2)}
  $
& 0.127357~(3)
\\ \\
\end{tabular}
\label{table4}
\end{table}

\begin{table}
\caption{Simplified $\mu$-integrals for the diagrams in Fig.\ \ref{ee}
(continuation)}
\begin{tabular}{lll}
\\
$\delta\epsilon^{ee}_{\mu6}$ & $\biggl(\frac{\pi^2}{3} - \frac{7}{2} \biggr)
\ln^2{\frac{M}{m}}
+ \biggl(-\frac{8\pi^2}{9} + \frac{28}{3} \biggr)\ln{\frac{M}{m}}
+\frac{\pi^4}{18} - \frac{5\pi^2}{36} - \frac{14}{3}
$  & $\biggl(\frac{\pi^2}{3} - \frac{7}{2} \biggr)
\ln^2{\frac{M}{m}}$
\\ \\
& $+  \frac{1}{18} \bigl( 2\pi^2-21\bigr)
 \bigl( 8\ln{2}-3\bigr)
+ \frac{3}{2} \int_0^1 {dx} \int_0^x {dy}c_{6}
$  & $+ \biggl(-\frac{8\pi^2}{9} + \frac{28}{3} \biggr)
\ln{\frac{M}{m}}$
\\ \\
& $
\int_{0}^{\infty} {d k^2}
\biggl[\frac{k^4}{(k^2 + a^2)^2} - \frac{k^2}{k^2 + 4}\biggr]
\int_0^1 {dv}\frac{v^2(1-v^2/3)}{4 + k^2(1-v^2)}  $
& $ +\frac{\pi^4}{18} - \frac{5\pi^2}{36} - \frac{14}{3}$
\\ \\
& &-0.1944535~(6)
\\ \\ \tableline
\\
$\delta\epsilon^{ee}_{\mu7 }$   & $\biggl(-\frac{\pi^2}{3} + \frac{5}{2}
\biggr) \ln^2{\frac{M}{m}}
+ \biggl(\frac{8\pi^2}{9} - \frac{20}{3} \biggr)  \ln{\frac{M}{m}}
- \frac{\pi^4}{18} - \frac{\pi^2}{36} + \frac{10}{3} $
& $\biggl(-\frac{\pi^2}{3} + \frac{5}{2} \biggr)
 \ln^2{\frac{M}{m}}$
\\ \\
& $-\frac{1}{18} \bigl( 2\pi^2-15\bigr)
 \bigl( 8\ln{2}-3\bigr)
- \frac{3}{2} \int_0^1 {dx} \int_0^x {dy} b c_{7}
$
& $+ \biggl(\frac{8\pi^2}{9} - \frac{20}{3} \biggr)
 \ln{\frac{M}{m}}$
\\ \\
& $
\int_{0}^{\infty} {d k^2}
\biggl[\frac{k^4}{(k^2 + a^2)^2} - \frac{k^2}{k^2 + 4}\biggr]
\int_0^1 {dv}\frac{v^2(1-v^2/3)}{4 + k^2(1-v^2)}  $
& $- \frac{\pi^4}{18} - \frac{\pi^2}{36} + \frac{10}{3} $
\\ \\
& &-0.8472862~(4)
\\ \\
\end{tabular}
\end{table}

\begin{table}
\caption{$c$-integrals for the diagrams in Fig.\ \ref{mm}}
\begin{tabular}{lll}
\\
$\delta\epsilon^{\mu\mu}_{1}$& $
\int_0^1 {dx} \int_0^x {dy} c_1
\int_0^{\infty} {d k^2} \biggl(\frac{\partial}{\partial a^2}\biggr)^2
\frac{1}{\pi}\int_0^{\pi} {d \theta} \sin^2 {\theta}$
\\  \\
& $(2 - \cos^{2} {\theta} - \cos^{4} {\theta})\frac{k^2(k^2+a^2)}{(k^2 +
a^2)^2 + 4 b^2 k^2 \cos^2{\theta}}\int_0^1 {dv}
\frac{v^2(1-v^2/3)}{4 + k^2(1-v^2)}$ & 0.297870~(1)
\\ \\ \tableline
\\ $\delta\epsilon^{\mu\mu}_{2}$      &   $\int_0^1 {dx}
\int_0^x {dy} c_2\int_0^{\infty} {d k^2}\biggl(\frac{\partial}{\partial
a^2}\biggr)^2\frac{1}{\pi}
\int_0^{\pi} {d \theta}\sin^2 {\theta}
$
\\  \\
& $(2 + \cos^{2} {\theta})
\frac{k^4(k^2+a^2)}{(k^2 + a^2)^2 ~+~ 4 b^2 k^2 \cos^2{\theta}}
\int_0^1 {dv}\frac{v^2(1-v^2/3)}{4 + k^2(1-v^2)}  $  &-0.759522~(5)
\\ \\ \tableline
\\
$\delta\epsilon^{\mu\mu}_{3}$     &   $2\int_0^1 {dx} \int_0^x {dy}
c_3\int_0^{\infty} {d k^2} \frac{\partial}{\partial a^2}
\frac{1}{\pi}\int_0^{\pi} {d \theta} \sin^2 {\theta}  $
\\ \\
& $(2 + \cos^2{\theta})
\frac{k^2(k^2+a^2)}{(k^2 + a^2)^2+4 b^2 k^2 \cos^2{\theta}}
\int_0^1 {dv}\frac{v^2(1-v^2/3)}{4 + k^2(1-v^2)}$ &-1.11839~(2)
\\ \\ \tableline
\\
$\delta\epsilon^{\mu\mu}_{4}$   & $-8\int_0^1 {dx} \int_0^x {dy}~ bc_4
\int_0^{\infty} {d k^2}\frac{\partial}{\partial a^2}
\frac{1}{\pi}\int_0^{\pi} {d \theta}\sin^2 {\theta} \cos^{2} {\theta}
$
\\ \\
& $(2+\cos^2{\theta})
\frac{k^2}{(k^2 + a^2)^2+4 b^2 k^2 \cos^2{\theta}}
\int_0^1 {dv}\frac{v^2(1-v^2/3)}{4 + k^2(1-v^2)} $ &-0.0484183~(2)
\\ \\ \tableline
\\
$\delta\epsilon^{\mu\mu}_{5}$   & $12\int_0^1 {dx} \int_0^x {dy}bc_5
\int_0^{\infty} {d k^2}\frac{\partial}{\partial a^2}
\frac{1}{\pi}\int_0^{\pi} {d \theta} \sin^2 {\theta}   $
\\ \\
& $\cos^{2} {\theta}
\frac{k^2}{(k^2 + a^2)^2 + 4 b^2 k^2 \cos^2{\theta}}
\int_0^1 {dv}\frac{v^2(1-v^2/3)}{4 + k^2(1-v^2)} $ &-0.0027928~(4)
\\ \\ \tableline
\\
$\delta\epsilon^{\mu\mu}_{6}$   & $-12
\int_0^1 {dx} \int_0^x {dy}c_6\int_0^{\infty} {d k^2}~
\frac{\partial}{\partial a^2} ~
\frac{1}{\pi}\int_0^{\pi} {d \theta} \sin^2 {\theta}     $
\\ \\
& $\cos^{2} {\theta}
\frac{k^2(k^2+a^2)}{(k^2 + a^2)^2 + 4 b^2 k^2 \cos^2{\theta}}
\int_0^1 {dv}\frac{v^2(1-v^2/3)}{4 + k^2(1-v^2)} $ &-0.0251636~(1)
\\ \\ \tableline
\\
$\delta\epsilon^{\mu\mu}_{7}$ & $-12\int_0^1 {dx} \int_0^x {dy}bc_7
\int_0^{\infty} {d k^2}\frac{1}{\pi}\int_0^{\pi} {d \theta}\sin^2 {\theta}    $
\\ \\
& $\cos^{2} {\theta}
\frac{k^2}{(k^2 + a^2)^2 + 4 b^2 k^2 \cos^2{\theta}}
\int_0^1 {dv}\frac{v^2(1-v^2/3)}{4 + k^2(1-v^2)}$ &-0.1453343~(4)
\\ \\
\end{tabular}
\label{table5}
\end{table}

\begin{table}
\caption{$c$-integrals for the diagrams in Fig.\ \ref{em}}
\begin{tabular}{lll}
\\
$\delta\epsilon^{\mu e}_{c1}$& $\frac{1}{3}
\int_0^1 {dx} \int_0^x {dy} c_1
\int_0^{\infty} {d k^2} \biggl(\frac{\partial}{\partial a^2}\biggr)^2
\frac{1}{\pi}\int_0^{\pi} {d \theta} \sin^2 {\theta}$
\\  \\
& $(2 - \cos^{2} {\theta} - \cos^{4} {\theta})\frac{k^2+a^2}{(k^2 +
a^2)^2 + 4 b^2 k^2 \cos^2{\theta}}\ln{k^2}
$ & -~15.~49349~(3)
\\ \\ \tableline
\\ $\delta\epsilon^{\mu e}_{c2}$      &   $\frac{1}{3} \int_0^1 {dx}
\int_0^x {dy} c_2\int_0^{\infty} {d k^2}\biggl(\frac{\partial}{\partial
a^2}\biggr)^2\frac{1}{\pi}
\int_0^{\pi} {d \theta}\sin^2 {\theta}
$
\\  \\
& $(2 + \cos^{2} {\theta})
\frac{k^2(k^2+a^2)}{(k^2 + a^2)^2 ~+~ 4 b^2 k^2 \cos^2{\theta}}
\ln{k^2}  $  &-1.235177~(4)
\\ \\ \tableline
\\
$\delta\epsilon^{\mu e}_{c3}$     &   $\frac{2}{3}\int_0^1 {dx} \int_0^x {dy}
c_3\int_0^{\infty} {d k^2} \frac{\partial}{\partial a^2}
\frac{1}{\pi}\int_0^{\pi} {d \theta} \sin^2 {\theta}  $
\\ \\
& $(2 + \cos^2{\theta})
\frac{k^2+a^2}{(k^2 + a^2)^2+4 b^2 k^2 \cos^2{\theta}}
\ln{k^2}$ &26.~74813~(3)
\\ \\ \tableline
\\
$\delta\epsilon^{\mu e}_{c4}$   & $-\frac{8}{3}\int_0^1 {dx} \int_0^x
{dy}~ bc_4 \int_0^{\infty} {d k^2}\frac{\partial}{\partial a^2}
\frac{1}{\pi}\int_0^{\pi} {d \theta}\sin^2 {\theta} \cos^{2} {\theta}
$
\\ \\
& $(2+\cos^2{\theta})
\frac{1}{(k^2 + a^2)^2+4 b^2 k^2 \cos^2{\theta}}
\ln{k^2} $ &1.304129~(5)
\\ \\ \tableline
\\
$\delta\epsilon^{\mu e}_{c5}$   & $4\int_0^1 {dx} \int_0^x {dy}bc_5
\int_0^{\infty} {d k^2}\frac{\partial}{\partial a^2}
\frac{1}{\pi}\int_0^{\pi} {d \theta} \sin^2 {\theta}   $
\\ \\
& $\cos^{2} {\theta}
\frac{1}{(k^2 + a^2)^2 + 4 b^2 k^2 \cos^2{\theta}}
\ln{k^2} $ &0.411544~(4)
\\ \\ \tableline
\\
$\delta\epsilon^{\mu e}_{c6}$   & $-4
\int_0^1 {dx} \int_0^x {dy}c_6\int_0^{\infty} {d k^2}~
\frac{\partial}{\partial a^2} ~
\frac{1}{\pi}\int_0^{\pi} {d \theta} \sin^2 {\theta}     $
\\ \\
& $\cos^{2} {\theta}
\frac{k^2+a^2}{(k^2 + a^2)^2 + 4 b^2 k^2 \cos^2{\theta}}
\ln{k^2} $ &-0.0503852~(6)
\\ \\ \tableline
\\
$\delta\epsilon^{\mu e}_{c7}$ & $-4\int_0^1 {dx} \int_0^x {dy}bc_7
\int_0^{\infty} {d k^2}\frac{1}{\pi}\int_0^{\pi} {d \theta}\sin^2 {\theta}    $
\\ \\
& $\cos^{2} {\theta}
\frac{1}{(k^2 + a^2)^2 + 4 b^2 k^2 \cos^2{\theta}}
\ln{k^2}$ &1.259721~(4)
\\ \\
\end{tabular}
\label{table6}
\end{table}

\end{document}